\documentclass[11pt]{article}
\usepackage{hyperref}
\pdfoutput=1
\begin{document}
\title{Dancing Droplets}
\author{Nate J Cira, Manu Prakash \\
\\\vspace{6pt} Stanford University, \\ Department of Bioengineering}
\maketitle
\begin{abstract}
Inspired by the observation of intricate and beautifully dynamic patterns generated by food coloring on clean glass slides, we have investigated the behavior of propylene glycol and water droplets on high energy surfaces. In this fluid dynamics video we show a range of interesting behaviors including long distance attraction, and chasing/fleeing upon contact. We present explanations for each of these behaviors including a mechanism for the long distance interactions based on vapor facilitated coupling. Finally we use our understanding to create several novel devices which: spontaneously align droplets, drive droplets in circles, cause droplets to bounce on a vertical surface, and passively sort droplets by surface tension. The simplicity of this system lends it particularly well to application as a toy model for physical systems with force fields and biological systems such as chemotaxis and motility. 
\end{abstract}

\end{document}